\documentclass[a4paper,fleqn,usenatbib]{mnras}

\usepackage{newtxtext,newtxmath}
\usepackage{hyperref}
\usepackage[T1]{fontenc}
\usepackage{ae,aecompl}
\usepackage{amsmath}
\usepackage[export]{adjustbox}
\newcommand*\diff{\mathop{}\!\mathrm{d}}

\usepackage{graphicx}	
\usepackage{amsmath}	
\usepackage{amssymb}	

\title[Rapidly Rotating Bipolytropic Structures]{A Numerical Method for Generating Rapidly Rotating Bipolytropic Structures in Equilibrium}

\author[ Kadam et al.]{
Kundan Kadam,$^{1}$\thanks{E-mail: kkadam1@lsu.edu}
Patrick M. Motl,$^{2}$
Juhan Frank,$^{1}$
Geoffrey C. Clayton$^{1}$ and  
\newauthor
Dominic C. Marcello$^{1}$
\\
$^{1}$Department\ of Physics \& Astronomy, Louisiana State University, Baton Rouge, LA 70803\\
$^{2}$The School of Sciences, Indiana University Kokomo, Kokomo, Indiana, 46904
}

\date{Accepted 2016 July 21. Received 2016 July 18; in original form 2016 April 12}
\pubyear{2016}

\begin{document}
\label{firstpage}
\pagerange{\pageref{firstpage}--\pageref{lastpage}}
\maketitle

\begin{abstract}
We demonstrate that rapidly rotating bipolytropic (composite polytropic) stars and toroidal disks can be obtained using Hachisu's self consistent field technique. The core and the envelope in such a structure can have different polytropic indices and also different average molecular weights. The models converge for high $T/|W|$ cases, where T is the kinetic energy and W is the gravitational energy of the system. 
The agreement between our numerical solutions with known analytical as well as previously calculated numerical results is excellent. 
We show that the uniform rotation lowers the maximum core mass fraction or the Sch$\ddot{\rm{o}}$nberg-Chandrasekhar limit for a bipolytropic sequence. We also discuss the applications of this method to magnetic braking in low mass stars with convective envelopes.
\end{abstract}




\section{Introduction}



Stellar models which use a polytropic equation of state (EoS), when combined with the structural and thermal equations, give us a reasonable understanding of the interiors of stars \citep{Chandrasekhar1939}. Even though the pressure is decoupled from the temperature, the self gravitating structures obtained by solving these equations show a remarkable similarity to the observations. Polytropes have been used to represent gaseous planets, main sequence stars, fully convective stars and even compact objects like white dwarfs and neutron stars \citep{HKT}. But for the stars which have a distinct core and envelope structure,
this approximation starts to break down. These stars with a well defined core (terminal age main sequence, subgiant, red giant, AGB etc.) behave in fundamentally different ways and we need a core-envelope structure in order to represent them faithfully. The stars on the main sequence also show a radiative core and convective envelope, or vice versa depending on their mass \citep{PadmanabhanV2}.  

One way to achieve a better approximation is to use bipolytropes (sometimes also referred to as composite polytropes), where the core and the envelope of a star are allowed to have different polytropic indices. This idea was first proposed by \cite{Milne1930} who suggested that the temperature, pressure and mass enclosed would be continuous at the core-envelope interface. 
\cite{Henrich1941}  implemented this condition for a star with an isothermal core and polytropic envelope to infer an upper limit in the core mass fraction.
\citet{SC1942} introduced a discontinuous jump in the molecular weights at the core-envelope boundary to deduce, the now famous, Sch$\ddot{\rm{o}}$nberg-Chandrasekhar (SC) limit. 
 Analytical solutions of single polytropic spheres are possible only in the cases where the polytropic index ($n$) equals zero (constant density), one or five, and only in the non-rotating and spherically symmetric scenario. Analytical solutions for bipolytropes are thus possible only in certain combinations of these three indices. For example, see \cite{Murphy1983} and \cite{Eggleton1998}.

Stellar rotation is the most important parameter affecting the behavior of a star after the initial mass and the metallicity \citep{Meynet2002}. 
Rotation has a profound effect on stellar structure equations, mass loss rate, angular momentum loss, mixing and stellar wind. The effects are also observed in the non-uniform surface temperature of stars and anisotropic structures in nebulae.
This is especially true for high mass stars as most are fast rotators with equatorial velocities only a factor of two less than breakup \citep{Kawaler1987}. 

In this paper we present the first computational method to determine the equilibrium structures of rotating bipolytropes. This is achieved by modifying Hachisu's self-consistent field (HSCF) method \citep{Hachisu1986a}; we call this technique the bipolytropic self-consistent field or the BSCF method for convenience. The core and envelope can have different polytropic indices and the composition difference between them can be represented as a ratio of the average molecular weights. We confirm the validity of this method by comparing the results with known analytical as well as numerical results. 
The bipolytropic models obtained can have a high degree of flatness or a high $T/|W|$, where $T$ is the kinetic energy and $W$ is the gravitational energy of the system.  
Although real stars rotate differentially \citep{Zeipel1925}, we consider only uniform rotation because this is a good first order estimate of the rotational properties. 

We show that when the core has sufficiently high average molecular weight as compared to rest of the star, a uniformly rotating star represented using a bipolytrope has a lower SC limit as compared to its non-rotating counterpart.
We then discuss applications of the BSCF method for the convective magnetic braking of low-mass stars. The bipolytropic models can also be used to probe internal structure and estimate rotational parameters of fast spinning gaseous planets and exoplanets.
The source code for the BSCF method is publicly
available via GitHub, so that others in the astrophysics community can make similar models for their own use. (Link to the GitHub repository: \href{https://github.com/kkadam/BSCF}{https://github.com/kkadam/BSCF}.)

\section{Bipolytropes}

The polytropic EoS assumes a power law relationship between the pressure of the gas and the density
\begin{equation}
 P=\kappa\rho^{1+\frac{1}{n}}= {\rm \kappa} \rho^{\gamma}
\end{equation}
where $n$ is the polytropic index, ${\rm \kappa}$ is the polytropic constant and $\gamma$ is the ratio of specific heats.
A bipolytropic star has one such polytropic index for the core ($n_{\rm c}$) and one for the envelope ($n_{\rm e}$). 
(The subscripts ${\rm c}$ and ${\rm e}$ correspond to the core and the envelope, and the subscript ${\rm i}$ corresponds to the core-envelope interface throughout this paper.)
In general, the polytropic constant is also different for the two regions. 
The physical quantities which are continuous across the core-envelope interface are pressure, temperature and the mass enclosed. 
While the pressure in the core and the envelope is represented by the respective polytropic EoS, at any point the total pressure must also be given by the sum of gas pressure and radiation pressure. Hence
\begin{equation}
P= \frac{ {\rm k_B}}{\mu {\rm m_H}} \rho T+\frac{1}{3} {\rm a} T^4
\end{equation}
where ${\rm k_B}$ is the Boltzmann constant, $\mu$ is molecular weight, ${\rm m_H}$ is the mass of the hydrogen atom, ${\rm a}$ is the radiation constant and $T$ is the temperature of the gas.
For an evolved star the core composition is different from that of the envelope, hence they have different molecular weights. 
If we assume temperature and pressure to be continuous across the boundary, then the density across this interface is not continuous, but is proportional to the ratio of molecular weights, 
\begin{equation}
\frac{\rho_{\rm ci}}{\rho_{\rm ei}}=\frac{\mu_{\rm c}}{\mu_{\rm e}}
\end{equation}
where $\rho_{\rm ci}$, $\rho_{\rm ei}$ are densities at the interface and $\mu_{\rm c}$ and $\mu_{\rm e}$ are the average molecular weights. These are the boundary conditions at the core-envelope interface of a bipolytrope. 
The ratio of molecular weights, 
\begin{equation}
\alpha=\frac{\mu_{\rm c}}{\mu_{\rm e}}
\end{equation}
represents the discontinuity at the core-envelope interface because of the difference in the composition between the core and the envelope.

At this point we introduce two key parameters that are used extensively in subsequent discussions of sequences of bipolytropic models.
The fractional core mass of a bipolytrope is defined as
\begin{equation}
\nu= \mathcal{M}_{\rm c}/ \mathcal{M}
\end{equation}
where $\mathcal{M}_{\rm c}$ is the mass of the core and $\mathcal{M}$ is the total mass. The fractional core radius is 
\begin{equation}
q= \mathcal{R}_{\rm c}/ \mathcal{R}
\end{equation}
where $\mathcal{R}_{\rm c}$ is equatorial radius of the core and $\mathcal R$ is equatorial radius of the whole configuration. The fractional core mass and radius are key to the discussions of model comparisons as well as any astrophysically relevant outcomes, as we shall see in sections \ref{NumericalTests} and \ref{NumericalResults}.

\section{The BSCF Method}
We can construct rapidly rotating single bipolytropic spheroidal as well as toroidal structures with the BSCF method.  
We make the following two assumptions while using the BSCF technique. 
The star (spheroid) or torus is assumed to be rotating axisymmetrically and the angular velocity ($\Omega$) throughout the structure is uniform.

\subsection{Equations}

In a frame of reference that is spinning with the star, the vector equation of motion of a self-gravitating body in equilibrium is given by, 
\begin{equation}
\frac{1}{\rho} \nabla P + \nabla (\Phi + \Omega^2 \Psi) =0
\end{equation}
where $\Psi$ is the coordinate-dependent part of the centrifugal potential and $\Phi$ is the gravitational potential.
With the specific enthalpy of the gas given by 
\begin{equation}
H= \int_0^{\rm P} {\frac{\diff{P}}{\rho}} ,
\end{equation}
Eq. 7 can be integrated to obtain
\begin{equation}
H+\Phi+\Omega^2 \Psi = {\rm C_B}
\end{equation}
with ${\rm C_B}$ as the integration constant -- analogous to the Bernoulli constant in classical fluid flows.
The last term is the centrifugal potential,
\begin{equation}
\Omega^2\Psi = -\Omega^2\frac{R^2}{2}
\end{equation}
where $R$ is the distance from the axis of rotation.

The outer boundary condition is that the density is zero at the stellar surface. Integrating Eq. 8 with a polytropic EoS gives the enthalpy as a function of density
 \begin{equation}
H = (1+n) {\rm \kappa} \rho^{\frac{1}{n}} {\rm, }
\end{equation}
hence the enthalpy is also zero at the outer boundary.

\subsection{Implementation}
\label{Implementation}

\begin{figure} 
\centering
\includegraphics[width=3in]{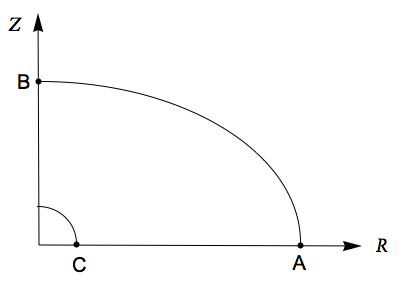}
\includegraphics[width=3in]{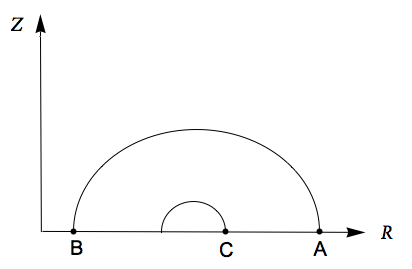}
\caption{A schematic of the meridional cross section of rotating figures in cylindrical coordinates, showing the boundary points are A, B and the core interface C, in the case of a spheroidal and toroidal bipolytropic structure respectively.}
\label{fig:schematic}
\end{figure}

In order to determine a bipolytropic equilibrium structure uniquely, a possible pair of parameters are the axis ratio (the ratio of the equatorial to the polar radius) and core radius fraction (the ratio of the equatorial radius of the core to the radius of the whole star).   
Fig. \ref{fig:schematic} shows a schematic diagram of a meridional cross section of a star (top panel) and of a toroidal configuration (bottom panel). 
Just like the HSCF method, we cannot specify the rotation rate (or the core mass fraction, in the case of a bipolytrope) a priori. We can only specify outer boundary points A and B, and as implemented here, the location of the core-envelope interface C.  
The lower the axis ratio, the faster the resulting configuration rotates, hence the angular velocity is controlled by means of points A and B. The core mass fraction is set via the location of point C. For a torus the points A and B specify an inner and an outer radius in the equatorial plane. 

In order to get an equilibrium solution, we solve Poisson's equation 
\begin{equation}
\nabla ^2 \Phi = - 4 {\rm \pi G} \rho
\end{equation}
and the hydrostatic equation (Eq. 9) iteratively as follows. We have to assume or ``guess" an initial density distribution satisfying the outer boundary conditions. The BSCF method is not sensitive to this initial distribution. 
A typical choice for the initial guess for a spheroid is a constant density cone, which would look like a right triangle joining points $\rm A$ and $\rm B$ in a meridional cross section. For a toroid, the cross section of initial density would look like a box. 
The potential field ($\Phi$) throughout the grid is calculated from this density distribution by solving Poisson's equation. 

We expect the integration constant for the core (${\rm C_{c}}$) to be different from that for the envelope (${\rm C_{e}}$) in general. Hence for the stellar core Eq.9 becomes
\begin{equation}
H+\Phi+\Omega^2 \Psi= {\rm C_{c}}   \quad {\rm when}  \quad \rho \geq \rho_{\rm ci}
\end{equation}
and the same equation for the envelope is
\begin{equation}
H+\Phi+\Omega^2 \Psi= {\rm C_{e}}    \quad {\rm when}  \quad \rho < \rho_{\rm ci}.
\end{equation}
Since $H=0$ at points A and B, we get
\begin{equation}
H_{\rm A}=-\Phi_{\rm A}-\Omega^2 \Psi_{\rm A}+ {\rm C_{e}} = 0,
\end{equation}
\begin{equation}
H_{\rm B}=-\Phi_{\rm B}-\Omega^2 \Psi_{\rm B}+ {\rm C_{e}} = 0.
\end{equation}
Hence we can obtain the angular velocity of the system
\begin{equation}
\Omega^2= -\frac{\Phi_{\rm A}-\Phi_{\rm B}}{\Psi_{\rm A}-\Psi_{\rm B}}.
\end{equation}
Once we have $\Omega$, we can obtain the integration constant for the envelope 
\begin{equation}
{\rm C_{e}}=\Phi_{\rm A}+\Omega^2\Psi_{\rm A}=\frac{\Phi_{\rm A} \Psi_{\rm B}-\Phi_{\rm B} \Psi_{\rm A}}{\Psi_{\rm B} - \Psi_{\rm A}}.
\end{equation}
Now from the continuity of pressure at the core-envelope interface, we can derive
\begin{equation}
H_{\rm ci}=H_{\rm ei} \frac{(n_{\rm c}+1)}{(n_{\rm e}+1)} \frac{\rho_{\rm ei}}{\rho_{\rm ci}},
\end{equation}
where $H_{\rm ei}$ can be obtained from ${\rm C_{e}}$
\begin{equation}
H_{\rm ei}=-\Phi_{\rm i}-\Omega^2 \Psi_{\rm i} + {\rm C_{e}}.
\end{equation}
The integration constant for the core can be obtained from
\begin{equation}
{\rm C_{c}}=H_{\rm ci}+\Phi_{\rm ci}+\Omega^2\Psi_{\rm i} .
\end{equation}
Now we can calculate the enthalpy throughout the star from Eqs. 13 and 14.
Once we have the enthalpy, we can calculate new density distributions of the core and the envelope from Eq. 11 as 
\begin{equation}
\frac{\rho_{\rm c}}{\rho_{\rm c-max}}= \frac{\rho_{\rm c}}{\rho_{\rm 0}}=\left( \frac{H_{\rm c}}{H_{\rm c-max}} \right)^{n_{\rm c}}
\end{equation}
and
\begin{equation}
\frac{\rho_{\rm e}}{\rho_{\rm e-max}}=\frac{\rho_{\rm e}}{\rho_{\rm ei}}=\left( \frac{H_{\rm e}}{H_{\rm e-max}} \right)^{n_{\rm e}}.
\end{equation}
We choose $\rho_{\rm c-max}$, or the central density $\rho_{\rm 0}$, as unity and the envelope density can be normalized with respect to this central density,
\begin{equation}
\frac{\rho_{\rm e}}{\rho_{\rm 0}}=\frac{\rho_{\rm ei}}{\rho_{\rm 0}}\left( \frac{H_{\rm e}}{H_{\rm e-max}} \right)^{n_{\rm e}}
=\frac{\rho_{\rm ci}}{\rho_{\rm 0}}  \frac{\mu_{\rm e}}{\mu_{\rm c}} \left( \frac{H_{\rm e}}{H_{\rm e-max}}\right)^{n_{\rm e}} ,
\end{equation}
and
\begin{equation}
\frac{\rho_{\rm e}}{\rho_{\rm 0}}=\frac{\mu_{\rm e}}{\mu_{\rm c}} \left(\frac{H_{\rm ci}}{H_{\rm c-max}} \right)^{n_{\rm c}} 
\left( \frac{H_{\rm e}}{H_{\rm e-max}}\right)^{n_{\rm e}} . 
\end{equation}
Thus we obtain a new density distribution for the next iteration cycle. 

The iteration process is repeated until the relative changes in $\rm C_{c}$, $\rm C_{e}$ and $\Omega^2$ are smaller than the prescribed convergence criterion, $\delta$. Here, we use $\delta =1 \times 10^{-3}$ for all our calculations, except for the models used for the Richardson extrapolation where $\delta =1 \times 10^{-4}$, in order to achieve better accuracy (see section \ref{SSB}).

Using the BSCF method as described above, we essentially choose the fractional core radius, $q$ as the control parameter.
The analytical relationship between fractional core mass, $\nu$ and fractional core radius, $q$ becomes double valued as the ratio of molecular weights, $\alpha$ increases and eventually at $\alpha = 3$, there cannot be equilibrium solutions above a certain value of $\nu$. This is the aforementioned SC limit for the core mass fraction (see Fig. \ref{fig:nuq}). 
Since we can only specify point $\rm C$ and hence the value of $q$ with the BSCF method, we cannot directly control the fractional core mass, $\nu$. For a given value of $q$, if there exist two equilibrium values of $\nu$, we observe that the method always converges to the solution with the lower $\nu$.
In order to overcome this drawback, we can modify the BSCF method to use the ratio of the density at the core side of the interface and the central density, $\rho_{ci}/ \rho_0$, as the control parameter instead of point $\rm C$, which uniquely determines the configuration. The equations involved are exactly the same as mentioned so far and the radius of the core ($q$) is calculated from the converged model. We discuss the implications of this modification of the BSCF method in sections \ref{SSB} and \ref{SCL}.

\subsection{Normalization}

We use the same normalization convention as used in \cite{Hachisu1986a}, mainly for the ease of comparison. The basis set of dimensionless quantities is maximum density (${\rm \rho_{0}}$), equatorial radius of the star ($\mathcal R$), and the gravitational constant (${\rm G}$). All computations are performed in the dimensionless form of the variables.

\section{Numerical Methods}

We use cylindrical geometry for our computational grid because a rotating body is naturally described in cylindrical coordinates.
The Poisson solver with this geometry was already available to us from our previous numerical work \citep{Motl2002}. This not only minimized the time for development of the BSCF method significantly, but also provided a well-tested, reliable and accurate method for our investigations.

\subsection{Poisson's equation}

The most compute-intensive section in the BSCF method is the Poisson solver, which is used for obtaining the gravitational potential from the specified density distribution at each integration time step.
For an assumed isolated density distribution, the boundary condition for the potential is that it goes to zero at infinity. 
But in practice, since the grid is always finite, we have to calculate the potential at the edge of the grid which is outside the mass distribution.  
We construct this boundary potential using a compact expression for the Green's function in cylindrical coordinates. As described in \cite{Tohline1999}, we get an analytical expression for the integral in terms of half-integer degree Legendre functions of the second kind. This means that we can obtain a very accurate solution for our discretized mass distribution. This technique is particularly useful for studying systems which conform well to azimuthal symmetry  because it can be applied to very flattened or elongated bodies without suffering penalties in either accuracy or computation time. 

We use an effective two dimensional version of the Poisson solver by setting the number of azimuthal modes supported in the Green's functions to one, thereby enforcing axial symmetry. In order to obtain the interior solution for the potential, we use an alternating direction implicit (ADI) scheme  \citep{Peaceman1955}.

\subsection{Discretization and Quality of Solutions} 

The computational grid geometry is cylindrical with azimuthal symmetry. The resolution of the grid is denoted by ${\rm NUMR} \times {\rm NUMZ}$, where ${\rm NUMR}$ is the number of cells in the $R$-direction and ${\rm NUMZ}$ is the number of cells in the vertical or $z$-direction. 

All physical quantities are calculated at the cell centers. The first cell in both $R$ and $z$-direction is a boundary cell and the unit distance between consecutive cells in both the directions is equal, i.e. $\diff{z}=\diff{R}$.

The quality of a converged solution is quantified using the virial error ($VE$), which is a global measure of the accuracy of the equilibrium of a non-linear dynamical system. 
\begin{equation}
VE=\frac{ \mid 2T+W+3\Pi \mid} { \mid W \mid}
\end{equation}
where $\Pi$ is the volume integral of the pressure,
\begin{equation}
\Pi = \int_V P \diff{V}
\end{equation}
This expression is derived from the scalar virial equation, hence the  virial error should ideally be zero for a system in equilibrium. 
A typical value for virial error with a resolution of $130 \times 130$ is $10^{-4}$. This value can be pushed lower by increasing the resolution, while simultaneously decreasing $\delta$ by an appropriate amount.

\section{Numerical Tests}
\label{NumericalTests}

In this section we compare the equilibrium configurations obtained through the BSCF technique with some known analytical as well as numerical results. We use a grid size of $130 \times 130$ for all our calculations, unless stated otherwise. The equatorial radius (location of the point A in Fig. \ref{fig:schematic}) of the star for all configurations is 120 cells. In the case of tori, this is the outer radius. 

\subsection{Polytropes, Maclaurin, and One-Ring Sequences}

We construct continuous polytropic structures by setting the polytropic index of the core equal to that of the envelope and the ratio of molecular weights to unity. These models should be identical to single polytropes and satisfy all the associated properties. 

\begin{figure}
\centering
\includegraphics[width=3.2in,right]{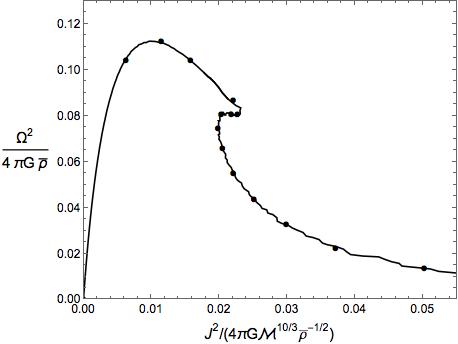}
\includegraphics[width=3.1in,right]{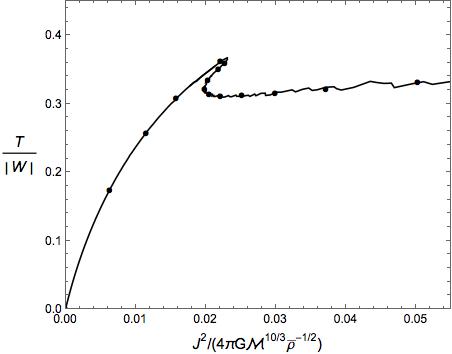}
\caption{Dimensionless squared angular velocity ($\Omega^2/ {\rm 4 \pi G} \bar{\rho}$), and rough stability indicator ($T/ | W|$) versus dimensionless squared angular momentum ($J^2/ {\rm 4 \pi G} \mathcal{M}^{10/3} \bar{\rho}^{-1/2} $) for the Maclaurin, Eriguchi-Sugimoto and Dyson-Wong sequences. The solid lines correspond to the models obtained using the BSCF method and the points are previously calculated values \citep{Hachisu1986a}.}
\label{fig:maclaurin}
\end{figure}

Maclaurin spheroids (MS) correspond to the equilibrium sequence of self-gravitating and uniformly rotating incompressible spheroids \citep{Tassoul1978}. 
This sequence transitions in a continuous fashion into Dyson-Wong (DW) toroids (or the one-ring sequence) through the Eriguchi-Sugimoto (ES) sequence, as the configurations rotate with increasing angular momentum \citep{Eriguchi1981}.
As a numerical test of the BSCF method, we reproduce the MS, ES and DW sequences for an incompressible fluid with $n=0$. The results are plotted in Fig. \ref{fig:maclaurin}. The smooth curves are obtained with the BSCF method as the point B moves along the z-axis (generating flatter spheroids by decreasing the axis ratio) and then along the R-axis (generating thinner toroidal structures) as depicted in Fig. \ref{fig:schematic}. 

We constructed similar sequences for other polytropic indices and these curves are also in excellent agreement with the calculations already performed (see Figs. 10 and 11 in \cite{Hachisu1986a}).

\subsection{Spherically Symmetric Bipolytropes}
\label{SSB}

As an additional test of the validity of the BSCF method, we constructed non-rotating bipolytropic models by setting the axis ratio to one. The properties of these structures were compared to analytical solutions.

\begin{figure} 
\centering
\includegraphics[width=3.2in]{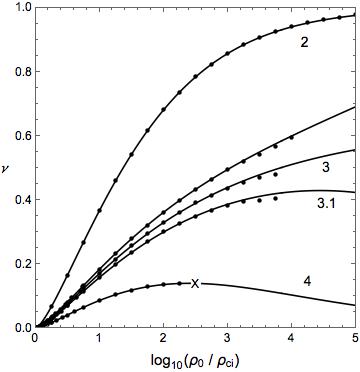}
\caption{Fractional core mass ($\nu$) versus logarithm of core density contrast ($\rho_{\rm 0}/ \rho_{\rm ci}$) for a $(n_{\rm c},n_{\rm e})=(5,1)$ bipolytrope with $\alpha=2$, $2.9$, $3$, $3.1$ and $4$. Points are BSCF results and the smooth curves are analytically determined \citep{Eggleton1998}. The cross denotes termination of the sequence in an SC limit (see Fig. \ref{fig:nuq}).}
\label{fig:egg_nu}
\end{figure}

\begin{figure} 
\centering
\includegraphics[width=3.3in]{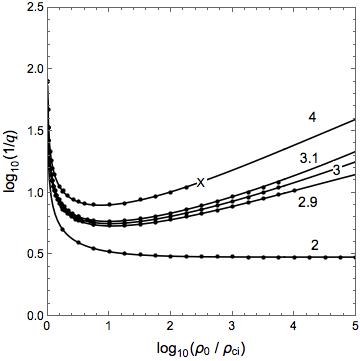}
\caption{
Logarithm of the inverse of fractional core radius ($q$) versus logarithm of core density contrast ($\rho_{\rm 0}/ \rho_{\rm ci}$) for a $(n_{\rm c},n_{\rm e})=(5,1)$ bipolytrope with $\alpha=2$, $2.9$, $3$, $3.1$ and $4$. Points are BSCF results and the smooth curves are analytically determined \citep{Eggleton1998}. 
The cross denotes termination of the sequence in an SC limit (see Fig. \ref{fig:nuq}).  
}
\label{fig:egg_q}
\end{figure}

The analytical case of spherically symmetric $(n_{\rm c},n_{\rm e})=(5,1)$ bipolytropes was first investigated by \cite{Eggleton1998}. They demonstrate that these bipolytropes exhibit an upper core mass fraction limit analogous to the SC
limit. We reproduce the analytical solutions with the BSCF method.  
In Figs. \ref{fig:egg_nu} and \ref{fig:egg_q}, respectively, we plot the core mass fraction ($\nu$) and  the logarithm of the inverse of core radius fraction ($q$) as a function of  the logarithm of the core density contrast, defined as $\rho_{\rm 0}/ \rho_{\rm ci}$. 
The points which are obtained from the BSCF models match the analytical sequences very well.

\begin{figure} 
\centering
\includegraphics[width=3.3in]{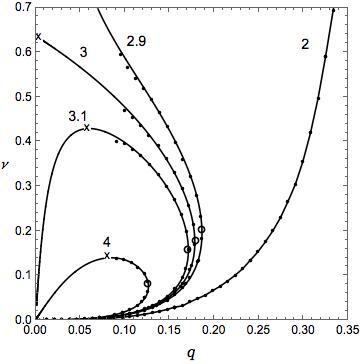}
\caption{Plot of fractional core mass $\nu$ versus fractional core radius $q$ for a bipolytrope with $(n_{\rm c},n_{\rm e})=(5,1)$ for different values of $\alpha$. 
The smooth curves are analytically determined and 
the points denote the sequences obtained using the BSCF method. The crosses are SC limits for a given $\alpha$ and open circles denote the switch from point $\rm C$ to $\rho_{ci}/ \rho_0$ as a free parameter in determining the mass of the core.}.
\label{fig:nuq}
\end{figure}

The aforementioned SC limit is better understood with the help of the $\nu-q$ curve. As depicted in Fig. \ref{fig:nuq}, for a high enough ratio of average molecular weight ($\alpha \geq 3$), the fractional mass shows an upper limit above which no equilibrium configurations exist. 
We can obtain all configurations up to this SC limit with the BSCF method. The open circles in Fig. \ref{fig:nuq} denote the switch in the BSCF method from using point $\rm C$ to using the ratio of the value of the density at the core side of the interface and the central density ($\rho_{ci}/ \rho_0$, also equal to the inverse of the density contrast) as the control parameter, in determining the size of the core. 
The exact location of the core interface or $q$ is now calculated using the enthalpy field. The correction to the fractional core mass $\nu$ is obtained by adding the contribution from the corresponding shell to the core mass.

As the density contrast increases or $\rho_{\rm ci}/\rho_{\rm 0}$ gets progressively smaller, the core gets more centrally condensed. This results in an equilibrium structure with a tenuous envelope and a core resembling a delta function, which contains fewer and fewer points. In order to calculate parameters $\nu $ and $q$ accurately, we need to resolve this core with a high resolution model.  
We implemented the Richardson extrapolation \citep{Richardson1911} at two different resolutions (total equatorial resolution of 120 and 240 cells) in order to get better estimates of $\nu$ and $q$ in Figs. \ref{fig:egg_nu},  \ref{fig:egg_q} and \ref{fig:nuq}. This is because higher resolution models which could resolve the core were prohibitive in computing time without parallelizing the code. This approach is also ultimately limited by the resolution.
This can be seen in Figs. \ref{fig:egg_nu} and \ref{fig:nuq} for $\alpha = 2.9$, $3$ and $ 3.1 $. For these curves, the BSCF results match the analytical sequences only up to certain points.
Obtaining these full curves 
may require higher resolutions and higher order Richardson extrapolations.

\section{Numerical Results}
\label{NumericalResults}

In this section we consider two applications of the BSCF technique. 
The first pertains to the effects of rotation on the SC limit. We next consider  the effects of rotation on the radius of gyration of low mass stars, which in turn affects the timescale for magnetic braking.

\subsection{Sch$\ddot{\rm{o}}$nberg-Chandrasekhar Limit}
\label{SCL}
As a star evolves along the main sequence, it burns the hydrogen in the central region into helium. 
Near the end of the main sequence phase, an inert helium core is formed at the center and the hydrogen continues to burn around this core in a shell. This core is essentially isothermal in order for it to be in thermal equilibrium. If the gas is assumed ideal, because of the interplay between internal pressure and self gravity, the surface pressure of such a core decreases strongly with the core mass fraction. \cite{SC1942} calculated that for a star with an isothermal core, a radiative envelope and a large enough molecular weight difference between the two,  
this core mass fraction has an upper limit above which the core cannot support the envelope. 
When this SC limit is reached, the star must respond by readjusting itself on a Kelvin-Helmholtz timescale by contraction of the core. The resultant liberation of the gravitational energy causes an increase in the luminosity. 
\cite{Applegate1988} argues that this exceeds the maximum luminosity that the radiative envelope can carry and the envelope density is forced down. This causes the envelope to expand and we get evolution along the red giant branch. 
This would be a first order approximate answer to the question ``Why stars become red giants". In real stars the process is much more complicated and non-linear, with nuclear burning, opacities, convection, chemical composition and various other quantities affecting the structure of the star (e.g. see \cite{SugimotoFujimoto2000}).

\cite{Maeder1971} was the first to discuss the effects of rotation on the SC limit.  He used first order perturbation analysis on the virial expression of a star with an isothermal core and a radiative envelope. In conclusion, the rotation could increase or decrease the SC limit, depending on the actual rotational profile in the stellar interior. For a uniform rotation, the SC limit decreases and even at the maximum rotation the decrease is only about 3\%.

\begin{figure} 
\centering
\includegraphics[width=3.3in]{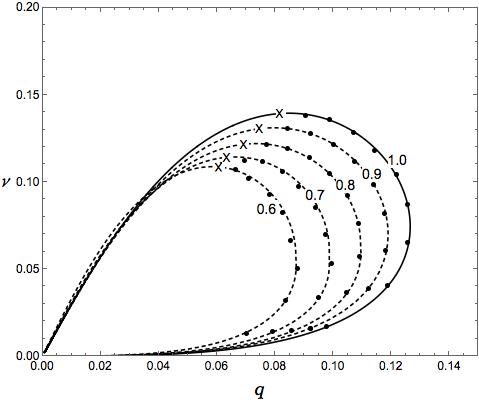}
\caption{Plot of fractional core mass $\nu$ versus fractional core radius $q$ for a bipolytrope with $(n_{\rm c},n_{\rm e})=(5,1)$ and $\mu=4$ for different values of axis ratios. 
The points denote the sequences obtained using the BSCF method. A cross denotes the Sch$\ddot{\rm{o}}$nberg-Chandrasekhar limit and the dashed lines represent the trend for the corresponding sequence.
The solid line is obtained from the non-rotating analytical solution.}
\label{fig:nuqrot}
\end{figure}

As we saw in the section \ref{SSB}, analytical bipolytropes also show an analogous SC limit above a certain jump in the average molecular weight. 
Here we observe the effect of rotation on the SC limit on a bipolytrope with an $(n_{\rm c},n_{\rm e})=(5,1)$  and a molecular weight ratio of $4$. 
Fig. \ref{fig:nuqrot} shows the $\nu-q$ sequences of these bipolytropes at different axis ratios. 
We see that 
as a bipolytrope rotates more rapidly, the maximum mass limit (denoted by a cross) decreases. A bipolytropic star rotating with an axis ratio of $0.6$ would encounter the SC limit at a mass fraction more than $20 \%$ lower than its non-rotating counterpart. 
Note that these values are quantitatively true only in the case of a bipolytropic structure with $(n_{\rm c},n_{\rm e})=(5,1)$, $\alpha=4$ and uniform rotation. 
This decrease is about an order of magnitude larger than that predicted by \cite{Maeder1971} for an isothermal core and a radiative envelope.
An SCF technique incorporating more realistic EoS and rotational profiles may be necessary in order to obtain a better estimate of the effects of rotation on the SC limit of real stars.

The change in the SC limit with rotation can have significant consequences for our understanding of stellar evolution, particularly for high mass stars as they do not have a convective envelope to brake their rotation efficiently. These stars also live on the main sequence for much shorter time so that they can still be rotating rapidly as they evolve off the main sequence. 
The SC limit determines the combustible stellar mass fraction of a star, thus affecting its lifetime spent on the main sequence. 
The effect due to rotation would be the next leading order of correction to the basic structure and evolution as would be dictated by the star's mass and chemical composition. 

Dependance of the SC limit on rotation may also result in an observational test of the rotational profiles in the stellar interiors. The slight changes in the frequency or position of stars on the HR diagram due to the shift in the SC limit can relate to the rotation law.
The SC limit should also be different in a close binary system as compared to an isolated star because of the rotational flattening. 
Sophisticated stellar evolution codes like MESA \citep{Mesa2011} are required to estimate the secular consequences of these results.

\subsection{Radius of Gyration}

Main sequence stars having masses between about $ 0.3 - 1.3 M_{\odot}$ have a radiative core and a convective envelope. These low mass stars can be represented as a bipolytrope with $(n_{\rm c},n_{\rm e})=(3,3/2)$ \citep{Rappaport1983, Beech1988a}. 
\cite{Rucinski1988} computed sequences of fractional radii of gyration of such spherically symmetric and rigidly rotating bipolytropic models.
The fractional radii of gyration for the core ($k_{\rm c}$) and envelope ($k_{\rm e}$) are defined such that the total angular momentum $J$ is
 \begin{equation}
 J = \mathcal{M} \mathcal{R}^2 \Omega  ( {k_{\rm c}}^2 + {k_{\rm e}}^2 ) 
 \end{equation} 
where $\mathcal{M}$ and $\mathcal{R}$ are the total mass and equatorial radius of the star respectively. This also means that the total radius of gyration is ${k_{\rm t}}^2={k_{\rm c}}^2+{k_{\rm e}}^2$. For an arbitrary stellar density distribution with the core-envelope interface located at $i$, we can calculate radii of gyration by integrating the moments of inertia,
 \begin{equation}
 {k_{\rm c}}^2 = \frac{1}{\mathcal{M} \mathcal{R}^2} \int_0^{\rm i} R^2 \diff{m},
 \end{equation} 
 and
 \begin{equation}
 {k_{\rm e}}^2 = \frac{1}{\mathcal{M} \mathcal{R}^2} \int_{\rm i}^{\mathcal{R}} R^2 \diff{m}.
 \end{equation} 
 
\begin{figure}
\centering
\includegraphics[width=3.2in]{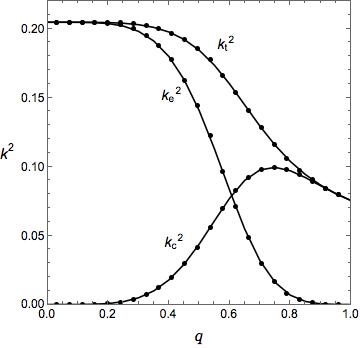}
\caption{Fractional radii of gyration for a non-rotating polytrope with $(n_{\rm c},n_{\rm e})=(3,3/2)$ and $\alpha=1$.  
The points are BSCF results and the smooth curves are calculated by 
\protect \cite{Rucinski1988}. 
${k_{\rm t}}^2$, ${k_{\rm c}}^2$ and ${k_{\rm e}}^2$
denote the curves for total, core, and envelope respectively.}  
\label{fig:rucinski}
\end{figure}

We recreate the sequences of radii of gyration in \cite{Rucinski1988} for the case of $(n_{\rm c},n_{\rm e})=(3,3/2)$ as a function of the core radius fraction. The original sequences correspond to numerical integrations of the expressions for $k_{\rm c}$ and $k_{\rm e}$, which are derived from a bipolytropic treatment of the spherically symmetric Lane-Emden equation without a discontinuity. 
The BSCF models constructed also have no discontinuity in molecular weight and the axis ratio is unity, hence they are spherically symmetric. As depicted in Fig. \ref{fig:rucinski} the results of the BSCF method are indistinguishable from the previously calculated curves.

\begin{figure} 
\centering
\includegraphics[width=3.2in]{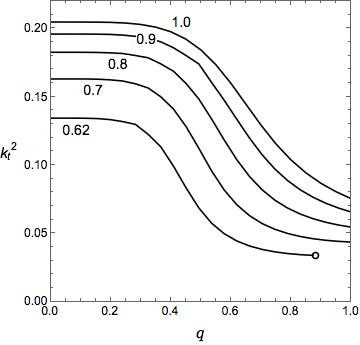}
\caption{Total radii of gyration for rotating bipolytropes with $(n_{\rm c},n_{\rm e})=(3,3/2)$ and $\alpha=1$.  The numbers denote the axis ratio of the models. The open circle denotes critical rotation.}
\label{fig:kt2}
\end{figure}

\begin{figure} 
\centering
\includegraphics[width=3.2in]{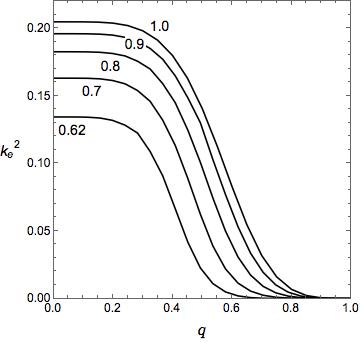}
\caption{Fractional radii of gyration of the envelope for rotating bipolytropes with $(n_{\rm c},n_{\rm e})=(3,3/2)$ and $\alpha=1$.  The numbers denote the axis ratio of the models.}
\label{fig:ke2}
\end{figure}

\begin{figure} 
\centering
\includegraphics[width=3.2in]{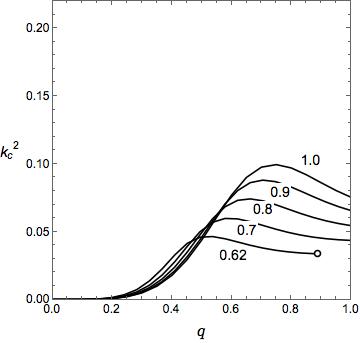}
\caption{Fractional radii of gyration of the core for rotating bipolytropes with $(n_{\rm c},n_{\rm e})=(3,3/2)$ and $\alpha=1$.  The numbers denote the axis ratio of the models. The open circle denotes critical rotation.}
\label{fig:kc2}
\end{figure}

With the BSCF method we can explore the effect of rotation on the radii of gyration.
Figs. \ref{fig:kt2}, \ref{fig:ke2} and \ref{fig:kc2} summarize the results as we progressively decrease the axis ratio. 
As a star rotates faster at a given fractional core radius, the corresponding radius of gyration steadily decreases. For the core, $k_{\rm c}^2$ shows a shift in the maximum value as the rotation increases. The sequence with an axis ratio of 0.62 terminates at $q=0.87$ (open circle in Figs. \ref{fig:kt2}, \ref{fig:kc2}) because this is the mass shedding limit. This means that the star is rotating at the critical velocity and no equilibrium configurations are possible above this value of $q$.

The dependence of the fractional radii of gyration on rotation can have implications for the rotational evolution of low-mass main sequence stars. 
A star can lose angular momentum through the process of magnetic braking as its wind is magnetically coupled to the rotation \citep{Mestel1968}.
This is an efficient mechanism for the loss of angular momentum for a star possessing a magnetic field. \cite{Skumanich1972} observed that G-dwarf stars in open clusters spin down following the empirical relation
\begin{equation}
v_{\rm e} \propto t^{-0.5}
\end{equation}
where $v_{\rm e}$ is equatorial velocity and $t$ is the age.
\cite{Vilhu1982} gives the semi-empirical law for magnetic braking  
\begin{equation}
\frac{\diff{J}}{\diff{t}} \propto \Omega^{N}
\end{equation}
where ${N}$ is the angular momentum loss parameter  (${N}=3$ for the Skumanich Law).
Here the underlying assumption is that the braking is not caused by the aging itself but the mechanism directly relates to the rotation of the star. 
Combining Eqs. 28 and 32, the characteristic timescale for magnetic braking can be given by
 \begin{equation}
    T_{\rm b} \approx J/\dot{J} \propto {k_{\rm t}}^2 \mathcal{M} \mathcal{R}^2 \Omega^{1-{N}}.
 \end{equation} 
 
For low-mass stars it is possible that the convective envelope decouples from the radiative core and the measurements of the magnetic braking apply to the envelope only \citep{Vilhu1986}. The characteristic timescale for magnetic braking then depends on the radius of gyration of the envelope alone, hence
 \begin{equation}
    T_{\rm b} \ \propto {k_{\rm e}}^2 \mathcal{M} \mathcal{R}^2 \Omega^{1-{N}}.
 \end{equation}

In order to calculate the timescale of braking of the convective envelope as a function of stellar mass, we follow the same recipe as in \cite{Rucinski1988}.
He used VandenBerg stellar models which have been calculated for a solar composition of ${\rm Y}=0.27$ and ${\rm Z}=0.0169$. The only input from these models that we use, apart from the overall mass-radius relationship, is the value of core mass fraction ($\nu$) for a given stellar mass.
The value of the angular momentum loss parameter decreases for more rapidly spinning stars \citep{Vilhu1982} and we assume ${N}=1$ for our calculations. 
Thus using the $k_e$ values obtained from the BSCF method and Table V in \cite{Rucinski1988} we can estimate the timescale for braking for the sequence of stars in the mass range $\mathcal{M} =0.5 - 1.2 M_{\odot}$.

 \begin{figure} 
\centering
\includegraphics[width=3in]{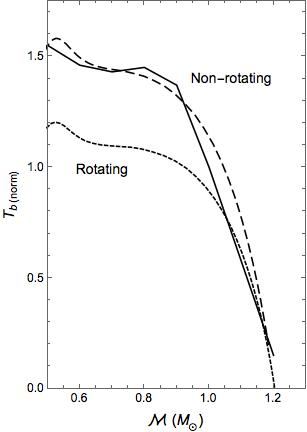}
\caption{The timescale for braking of the convective envelope ($T_{\rm b (norm)}$) as a function of stellar mass ($\mathcal{M}$). The solid line corresponds to the calculations 
performed by \protect \cite{Rucinski1988} using non-rotating VandenBerg models. The dashed lines correspond to the non-rotating and 
rotating bipolytropic sequences generated using the BSCF method for the same models. $T_{\rm b(norm)}$ is normalized to the value at $1 M_{\odot}$ for the Rucinski sequence.}
\label{fig:braking}
\end{figure}

Fig. \ref{fig:braking} summarizes the effects of rotation on the braking timescale of the convective envelope. For ease of comparison, the braking timescale is normalized to the value at $1 M_{\odot}$ for the \cite{Rucinski1988} sequence (solid black line in Fig. \ref{fig:braking}).
The timescales for non-rotating bipolytropes calculated using the BSCF method are consistent with those estimated by Rucinski. 
Notice that the timescale plateaus and shows little variation below $0.9 M_{\odot}$ but in the mass range of $0.9 - 1.2 M_{\odot}$, it drops precipitously.   
For rotating stars with an axis ratio of 0.7, this plateau below $0.9 M_{\odot}$ occurs at a significantly (about 25\%) lower characteristic braking timescale. 
We chose an axis ratio of 0.7 in order to demonstrate the most extreme difference that rotation could reasonably make,
since this axis ratio puts the rotational velocity of a star close to breakup.
The sequences of braking timescales for axis ratios other than 0.7 will show trends similar to Fig. \ref{fig:ke2}.

This analysis may exaggerate the difference in braking time because as the stars slow down, the evolution of the fractional radius of gyration is not considered ($k_{\rm e}$ is assumed constant throughout the above exercise). We show that the convective envelopes of these low-mass stars contain less initial angular momentum than previously estimated, when we take the distortion due to the rotation into account, irrespective of subsequent evolution. 
This should be reflected in the estimations of the braking time and all previous such calculations can be considered as upper limits. 
These results affect the observability of rapidly rotating stars in a given cluster as fast rotating stars brake on a shorter timescale.
The results could also have implications for the evolution of short period binary systems, because the same braking scheme is generally used for estimating angular momentum loss in close binary systems \citep{Vilhu1982}.

\section{Conclusions }
We present a numerical method to obtain self- consistent rapidly rotating bipolytropic models in equilibrium. This is achieved using the BSCF method. The sequences generated with this method agree with previously calculated and well established analytical as well as numerical results. 
These models can be used to investigate the effects of rotation on the internal structures and properties of stars. We show that the structural changes accompanying rotation lower the Sch$\ddot{\rm{o}}$nberg-Chandrasekhar mass limit in the case of uniform rotation. The rotational evolution of low-mass main sequence stars is also affected when we consider self-consistent rotating stellar models.

Rotating bipolytropes can also prove useful in the investigations into the internal structure of gaseous planets. Using a bipolytropic model similar to \cite{Criss2015}, we calculated the orbital period, normalized moment of inertia and the zonal harmonic coefficients for planet Jupiter with reasonable accuracy. Similar rotating models can be used to narrow down the internal structures of fast spinning exoplanets (for example, $\beta$ Pictoris b  \citep{Snellen2014}).

In order to further improve the models generated using the BSCF method, differentially rotating configurations with constant rotational velocity ($v$-constant) or constant angular momentum per unit mass ($j$-constant) could be obtained by following the recipe in \cite{Hachisu1986a}. 
This technique may also be used in order to generate rotating multi-polytropic models with an onion-like structure.
We believe that the BSCF method is a step towards more realistic, yet relatively simple astrophysical models which minimize the complexity of the input physics and help us gain an intuitive understanding of the results obtained.  

\section*{Acknowledgements}
We have greatly benefitted from discussions of this topic with Prof. Joel E. Tohline as well as from technical derivations related to the structure and stability of bipolytropes that he has provided in online documents (search, for example, ``Tohline Bipolytropes"). We would also like to thank the anonymous referee whose insightful comments led to the use of $\rho_{ci}/ \rho_0$ as a free parameter in the BSCF method.

We wish to acknowledge the support from the National Science Foundation through CREATIV grant AST-1240655. The numerical work was carried out using the computational resources of the Louisiana Optical Network Initiative (LONI). 




\bibliographystyle{mnras}
\bibliography{references_file} 








\bsp	
\label{lastpage}
\end{document}